\documentclass{article}

\usepackage{arxiv}

\usepackage[utf8]{inputenc} % allow utf-8 input
\usepackage[T1]{fontenc}    % use 8-bit T1 fonts
\usepackage{hyperref}       % hyperlinks
\usepackage{url}            % simple URL typesetting
\usepackage{booktabs}       % professional-quality tables
\usepackage{amsfonts}       % blackboard math symbols
\usepackage{nicefrac}       % compact symbols for 1/2, etc.
\usepackage{microtype}      % microtypography
\usepackage{lipsum}
\usepackage{graphicx}
\graphicspath{ {./images/} }
\usepackage{soul}
\usepackage{float}
\usepackage{graphicx}
\usepackage{caption}
\usepackage{subcaption}
\usepackage{amsmath}
\usepackage{multirow}
\usepackage{enumitem}
\usepackage{mathtools}
\DeclarePairedDelimiter\floor{\lfloor}{\rfloor}

\title{Predicting Elastic Properties of Materials from  Electronic Charge Density Using 3D Deep Convolutional Neural Networks}

\author{
 Yong Zhao $^{1}$,Kunpeng Yuan$^{2,3}$, Yinqiao Liu $^{2,4}$,Steph-Yves Louis$^{1}$, Ming Hu $^{2,*}$, and Jianjun Hu $^{1,*}$\\
  $^1$Department of Computer Science and Engineering\\
  University of South Carolina\\
  Columbia, SC, 29201\\
   * Correspondence:\texttt{jianjunh@cse.sc.edu (J.H.)}
\And
  \\
  $^2$Department of Mechanical Engineering\\
  University of South Carolina\\
  Columbia, SC, 29201\\
  * Correspondence:\texttt{hu@sc.edu (M.H.)}
\And
\\
$^3$Key Laboratory of Ocean Energy Utilization and Energy Conservation of Ministry of Education\\
School of Energy and Power Engineering,Dalian University of Technology\\
Dalian, 116024, China\\
\And
  \\
  $^4$Key Laboratory of Materials Modification by Laser, Ion and Electron Beams\\
  Dalian University of Technology), Ministry of Education\\
  Dalian, 116024, China\\
}

\begin{document}
\maketitle
\begin{abstract}
Materials representation plays a key role in machine learning based prediction of materials properties and new materials discovery. Currently both graph and 3D voxel representation methods are based on the heterogeneous elements of the crystal structures. Here, we propose to use electronic charge density (ECD) as a generic unified 3D descriptor for materials property prediction with the advantage of possessing close relation with the physical and chemical properties of materials. We developed an ECD based 3D convolutional neural networks (CNNs) for predicting elastic properties of materials, in which CNNs can learn effective hierarchical features with multiple convolving and pooling operations. Extensive benchmark experiments over 2,170 Fm$\bar{3}$m face-centered-cubic (FCC) materials show that our ECD based CNNs can achieve good performance for elasticity prediction. Especially, our CNN models based on the fusion of elemental Magpie features and ECD descriptors achieved the best 5-fold cross-validation performance. More importantly, we showed that our ECD based CNN models can achieve significantly better extrapolation performance when evaluated over non-redundant datasets where there are few neighbor training samples around test samples. As additional validation, we evaluated the predictive performance of our models on 329 materials of space group Fm$\bar{3}$m by comparing to DFT calculated values, which shows better prediction power of our model for bulk modulus than shear modulus. Due to the unified representation power of ECD, it is expected that our ECD based CNN approach can also be applied to predict other physical and chemical properties of crystalline materials.\end{abstract}

% \keyword{graph convolutional neural networks; global attention mechanism; deep learning; materials property prediction; OQMD; materials project}

% keywords can be removed
%\keywords{First keyword \and Second keyword \and More}

\section{Introduction}
Due to its time and cost efficiency, data-driven machine learning approaches have been increasingly used for material property prediction~\cite{cao2019convolutional,olsthoorn2019band} and materials screening and discovery ~\cite{sendek2018machine,avery2019predicting}. Although the great potential of machine learning in material discovery is widely acknowledged, it has yet to achieve high success as it has in other scientific fields. There are two major challenges to address to realize their potential. The first one is that in materials science there are usually only a small amount of characterization/property data (labelled samples) available, the so-called small data set problem \cite{cubuk2019screening}. For example, the number of materials with characterized thermal conductivity are less than 400 \cite{chen2019machine} while the number of materials with characterized ionic conductivity are even less than 50 \cite{sendek2018machine}. With limited data, a major challenge for building an accurate prediction model for a target material property is how to find suitable materials descriptors, which is a key factor that determines the prediction performance of machine learning models. A descriptor encodes materials' elemental, structural, and other physical information into a representation that machine learning algorithms can map to materials properties~\cite{cecen2018material,ward2016general,kajita2017universal,de2016comparing}. 

In the past decade, a large number of descriptors have been proposed to encode materials ~\cite{rupp2012fast,cecen2018material,ward2016general,villars2004data,faber2015crystal,rupp2015machine,kajita2017universal,bartok2013representing,szlachta2014accuracy,de2016comparing,ghiringhelli2015big,meredig2014combinatorial,pham2017machine}. In general, those descriptors are based on materials composition, their electronic or geometric structures as shown by the integrated feature calculation routines as implemented in the Matminer package \cite{ward2018matminer}. A widely used set of material composition features is the Magpie features, which are based on the statistics of elemental properties in a material \cite{ward2016general}. Mendeleev numbers (MN) has also been used by P. Villars et al.~\cite{villars2004data}  to classify  chemical systems by using the minimum and maximum MN versus the ratio between the minimum and the maximum MV. Ghiringhelli et al.~\cite{ghiringhelli2015big} developed 23 primary features, based on atomic properties, to explore the energy difference of zinc blende, wurtzite, and rocksalt semiconductors. Logan Ward et al.~\cite{ward2016general} presented a comprehensive set of features for a wide variety of material compositions. This set contains four unique categories: stoichiometric attributes, elemental property statistics, electronic structure attributes, and ionic compound attributes. Elemental descriptors have achieved great success in predicting band gaps \cite{zhuo2018predicting}, formation energies\cite{jha2019enhancing}, crystal system\cite{}, and etc.  But these descriptors have their severe limitations: elemental descriptors are merely based on material compositions while most materials properties are strongly dependent on their atomic structures. There are also materials that share the same composition but exist in completely different structures. It is a common understanding that the most important information for analyzing a material's property is its structure. How atoms coordinate and interact with each other conveys rich information on the properties of the materials. Therefore, structural features play a key role in developing prediction models of materials. Currently, there are several successful applications that use structural features to predict materials properties~\cite{cecen2018material,faber2015crystal,rupp2015machine,kajita2017universal,bartok2013representing,szlachta2014accuracy,de2016comparing,pham2017machine}. Rupp and colleagues applied the Coulomb matrix (CM) features for predicting the atomization energies of small isolated organic molecules ~\cite{rupp2012fast,faber2015crystal,rupp2015machine}. CM formulates the electrostatic interaction between nuclei into a matrix representation. Pham et al.~\cite{pham2017machine} developed the orbital-field matrix (OFM) descriptor, based on the distribution of valence shell electrons, to predict formation energies and atomization energies with high accuracy. Bart{\'o}k et al. ~\cite{bartok2013representing} proposed the Smooth overlap of atomic positions (SOAP), which describes the similarity between two atomic environments to define a metric in the structural cell. The local similarity can be combined further to form a global measure of similarity for the evaluation of molecular properties~\cite{de2016comparing}. More recently, voxel grid representation with atom features has been proposed to predict Hartree energies\cite{kajita2017universal}. Atom density and related continuous representations have also been proposed for materials representation and are used for crystal structure generation\cite{noh2019inverse,willatt2019atom}. Graph neural networks have also been introduced to learn structural representation from material structures for predicting formation energy, band gaps, bulk modulus and etc with great success \cite{xie2018crystal,chen2019graph}. On the other hand, deep learning has been utilized to extract three dimensional (3D) spatial features for material property prediction. In ~\cite{cecen2018material,kajita2017universal}, 3D CNNs have been applied to extract 3D geometric features from material microstructures represented as 3D matrices ~\cite{cecen2018material}. In this work, a dataset with 5,900 microstructures was created, where a microstructure is the quantification of the material structure. Each microstructure is represented by a feature matrix of dimension $51\times 51\times 51$, where each feature corresponds to a vector. Kajita et al.~\cite{kajita2017universal} developed a descriptor called Reciprocal 3D Voxel Space Descriptor (R3DVS) from the distributions of the valence electron density for 680 oxides. The authors enlarge the dataset by rotating R3DVS for testifying invariance of R3DVS to rotation and translation. R3DVS compacts the density of electrons in the bond generation.

In this paper, we propose to leverage convolutional neural networks (CNNs) to learn physically meaningful features from the three dimensional electronic charge density (ECD) of materials for elastic property prediction. Since physical and chemical properties of materials are related to the transferability between atoms (nuclei) and the presence of electronic charges or electronic multipoles on atoms or molecules~\cite{henkelman2006fast,ouyang2015competing,qin2017external,qin2018lone}, extraction of informative features from materials ECD can help predict materials properties. The ECD of a material can be calculated as a 3D matrix that describes the amount of electronic charge per volume. It represents the charge of electrons in the effective material space. By explicitly encoding the geometry of materials, ECD is supposed to have high transferability with respect to different compositions and structures \cite{gong2019predicting}.
As ECD captures both geometrical and electronic structural features, 3D distribution of electronic charge density would have the advantage over classical 1D and 2D descriptors as well as heterogeneous 3D structural descriptors in terms of the correlation with the electrochemical properties of materials. Indeed, ECD and its related electronic properties such as the electrostatic potential, electron localization function and non-covalent interaction index have been used to analyze many materials characteristics, including bonding, defects, stability,reactivity, and electron, ion and thermal transport \cite{gong2019predicting}. For example, ECD was used to predict 8 materials properties by using the Fourier coefficients of the planar averaged Kohn-Sham charge density fingerprint features \cite{pilania2013accelerating}. Abraham et al.~\cite{abraham2015electronic} calculated 2D charge density to predict the chemical bonding and charge transfer in magnetic compounds. However both approaches failed to take advantage of the flexibility of the 3D representation \cite{choi2019predicting}. Compared to conventional ML models, 3D CNNs can better link 3D descriptors to the properties efficiently as shown by~\cite{cecen2018material} (linkage between microstructure and homogenized property) and ~\cite{kajita2017universal} (linkage between R3DVS and Hartree energies, testify the invariance to rotation for R3DVS). We believe that the unified representation of ECD makes it easier to learn unified continuous representation for facilitating downstream prediction tasks by deep convolutional neural networks \cite{noh2019inverse}.

% \hl{Our intuition is that as a fundamental representation of electron interactions, xxx the ECD Two different strategies are applied to the ECD representations}.   

We explored two types of convolutional neural network models for ECD based elastic property prediction. One is the standard 3D CNNs with two convolutional layers. The other one is a projected 2D CNN models, in which the ECD matrix is converted to three different image-like representations which are then fed to three 2D CNN networks whose outputs are then fused together. This allows us to exploit the powerful hierarchical representation capability of 2D CNNs as shown in computer vision ~\cite{chen2017multi,simonyan2014very,maturana2015voxnet,szegedy2017inception,krizhevsky2012imagenet}. We then conducted extensive benchmark experiments based on a dataset consisting of 2170 Fcc structured materials and 11 non-redundant datasets generated by leaving one-element-out at a time. 

Our contributions can be summarized as follows:
\begin{itemize}
    \item We propose to exploit the ECD descriptor as unified 3D materials representation and combine it with two types of 3D CNNs for materials elastic property prediction. We also developed a fusion CNN model based on CNN+Magpie and CNN+ECD models.
    \item We developed a standard benchmark ECD dataset, named ``FCC2170'' calculated from 2710 Fcc Structured materials from ICSD. This database is characterized by its highly redundant samples with very similar compositions. We also developed 11 non-redundant datasets for evaluating the extrapolation capability of ECD+CNN models.
    \item We performed extensive prediction experiments over the aforementioned datasets using 5-fold cross validation. Our results show that our ECD+CNN can be complementary to elemental Magpie feature based models while can significantly outperform them over non-redundant datasets, thus demonstrating superior performance on some extrapolation experiments.
    \item We analyzed the situations when our ECD+CNN models perform better by visually inspecting the distribution of test samples and training samples in the 2D space mapped from the learned features.
    \item We validated the prediction performance of our models by comparison with DFT-calculated bulk and shear modulus for a set of 329 materials of the space group Fm$\bar{3}$m collected from  the Open Quantum Materials Database (OQMD) database.
\end{itemize}

% Our intention here is to introduce a new descriptor ECD and its combination with deep neural networks for predicting physical and chemical properties.  We detail how we use CNNs to explore the usefulness of ECD descriptor in the following sections.

%%%%%%%%%%%%%%%%%%%%%%%%%%%%%%%%%%%%%%%%%%
\section{Methods}
%%%%%%%%%%%%%%%%%%%%%%%%%%%%%%%%%%%%%%%%%%
\subsection{Datasets}
\label{section:dataset}

Here we discuss how we create the benchmark datasets for training and validating our proposed method. Due to the high computational cost to calculate electronic charge density for all the materials in the Materials Project database, we decide to focus only on materials of one specific space group. First we retrieved 2170 material structures of Fm$\bar{3}$m space group (excluding Lanthanides and Actinides) from the Materials Project (MP) database (\url{https://materialsproject.org}).  We chose the Fm$\bar{3}$m structure because its structure is simple and it takes less time to calculate the related elastic properties using DFT. Most materials of the Fm$\bar{3}$m space group do not have the charge density or elastic properties available in the MP database. Hence, we calculated both the electronic charge density~\cite{artmann1948berechnung} and the elastic property~\cite{wu2007crystal} using VASP~\cite{kresse1993ab,kresse1996efficiency,kresse1996efficient} for the 2,170 samples to form the ``FCC2170'' dataset . Table~\ref{t:num_elem} lists top 11 elements that are contained in at least 200 materials of our FCC2070 dataset. Among them, most are from Group 1 (Lithium, Sodium, Potassium, Rubidium, and Caesium), Group 13 (Indium and Thallium) and Group 17 (Fluorine, Chlorine, and Bromine). The rest includes Scandium from Group 3. With this dataset, we then use the commonly used cross-validation method to evaluate our model's interpolation performance as done in most machine learning based property prediction studies \cite{chen2019graph}. 

To validate our model's extrapolation capability, we define a set of leave-one-element-out datasets. For all samples in FCC2170, we first select those samples containing one specific element E as the test set, and then designate the remaining samples as the training set. These datasets are called FCC-E-N datasets, where E is the element of interest and N is the number of training samples without element E. Statistics of all these non-redundant datasets generated from FCC2170 for elements contained in more than 200 materials are shown in Table~\ref{t:num_elem}.

\begin{table}[H]
  \center
  \caption{Statistics of non-redundant Datasets}
  \begin{tabular}{|c | c | c | c | c | c | c |} 
  \hline
  Element & F & K & Rb & Cs & Na & Cl \\
  \hline
    dataset & FCC-F-1775 & FCC-K-1800 & FCC-Rb-1802 & FCC-Cs-1814 & FCC-Na-1877 & FCC-Cl-1880 \\
    \hline
  train set size & 1755 & 1800 & 1802 & 1814 & 1877 & 1880 \\
    \hline
    test set size & 415 & 370 & 368 & 356 & 293 & 290\\
    \hline
    \hline
  Element &In & Br & Li & Sc & Tl&-\\
  \hline
    dataset & FCC-In-1937 & FCC-Br-1938 & FCC-Li-2148 & FCC-Sc-1952 & FCC-Tl-1966&-\\
    \hline
  train set size & 1937 & 1938 & 2148 & 1952 & 1966 &-\\
    \hline
    test set size & 233 & 232 & 222 & 218 & 204&-\\
    \hline
  \end{tabular}
  \label{t:num_elem}
\end{table}

\subsection{Representations of Materials}
We studied and compared two material representations for elastic property prediction including Magpie~\cite{ward2016general} and electronic charge density(ECD)~\cite{silvi1994classification}. 

\begin{itemize} %[leftmargin=*,labelsep=5.8mm]
  \item Magpie features\\
    Magpie (Materials-Agnostic Platform for Informatics and Exploration) is an extensive set of features related to the constituent elements in materials. The set covers a broad range of physical and chemical properties that fall into four different categories: stoichiometric features, elemental property statistics, electronic structure features, and ionic compound features~\cite{ward2016general}. Stoichiometric features only contain the number of elements in the compound and their several $L^{p}$ norms~\cite{ward2016general}. Elemental property statistics are calculated by computing several statistics (e.g., average, minimum, maximum, range and mode) of 22 different elemental properties~\cite{ward2016general}. These properties include row and column on the periodic table, average atomic number, the range of atomic radii between all elements presenting in compositions, Mendeleev number, atomic weight, covalent radius, electro-negativity. Electronic structure features are the average fraction of s, p, d and f valence electrons~\cite{meredig2014combinatorial}. Ionic compound features include the capability of forming an ionic compound (when we assume all elements present in a single oxidation state) and two adaptions for calculating the fractions of a compound based on electronegativity~\cite{callister2007materials}.

  \item electronic charge density\\
ECD in the form of 3D structural matrix represents the spatial distribution of electronic charge density in crystalline materials. It can be calculated by local quantum-mechanical functions related to the Pauli exclusion principle~\cite{silvi1994classification}. The \textit{ab initio} method is used to calculate Hartree-Fock wavefunction and the electron localization function (ELF)~\cite{artmann1948berechnung}. A single determinant wave function is calculated on a grid in the 3D space by hartree Fock or Kohn Sham orbitals $\varphi_{i}$ as following:
  
    \begin{equation}
      ELF=\frac{1}{1+(\frac{D}{D_{h}})^{2}}
    \end{equation} 
    where 
    \begin{equation}
    \begin{split}
      &D=\frac{1}{2}\sum_{i}|\Delta\varphi_{i}|^{2}-\frac{1}{8}\frac{|\Delta\rho|^{2}}{\rho}\\
      &D_{h}=\frac{3}{10}(3\pi^{2})^{5/3}\rho^{5/3}
    \end{split}
    \end{equation}
    where $ELF$ has values between 0 and 1, where 1 means the perfect localization. Figure~\ref{fig:visual_ECD} shows the visualizations for the ECDs of six representative materials, namely $\mathrm{SRCaIn_{2}}$, $\mathrm{Mn_{23}C_{6}}$, $\mathrm{VSiOs_{2}}$, RbI, CsBr, and $\mathrm{Rb_{2}TeBr_{6}}$, where $\mathrm{SRCaIn_{2}}$, $\mathrm{Mn_{23}C_{6}}$, and $\mathrm{VSiOs_{2}}$ possess high bulk modulus. These visualizations consist of points that correspond to the values in a material's ECD matrix. The color and area of each point represents the size of each value and together show the distribution of a material's electron clouds. When the value of these points are plotted, we found that points appear in both thick and thin clouds, within the cubes, as shown in subfigures~\ref{fig:SrCaIn2},~\ref{fig:Mn23C6}, and~\ref{fig:VSiOs2}. Subfigures~\ref{fig:RbI},~\ref{fig:CsBr}, and~\ref{fig:Rb2TeBr6} show a clear difference from the top-row figures. In these figures, there are some empty spaces in the cubes and some dense clusters present in the remaining area. These observations correspond to the physical reality that materials with high bulk modulus usually have active electrons orbiting across the whole space strongly when compared to materials with lower bulk modulus. Among all six materials, we find that although the ECD visualizations share many similar characteristics, there are a few distinct differences between them. These minor variations make it possible for us to employ 3D CNNs to learn the structural and physical patterns that may characterize the material's elastic properties.

    \begin{figure}[H]
      \centering
        \begin{subfigure}[b]{.3\linewidth}
          \centering
          \includegraphics[width=\linewidth]{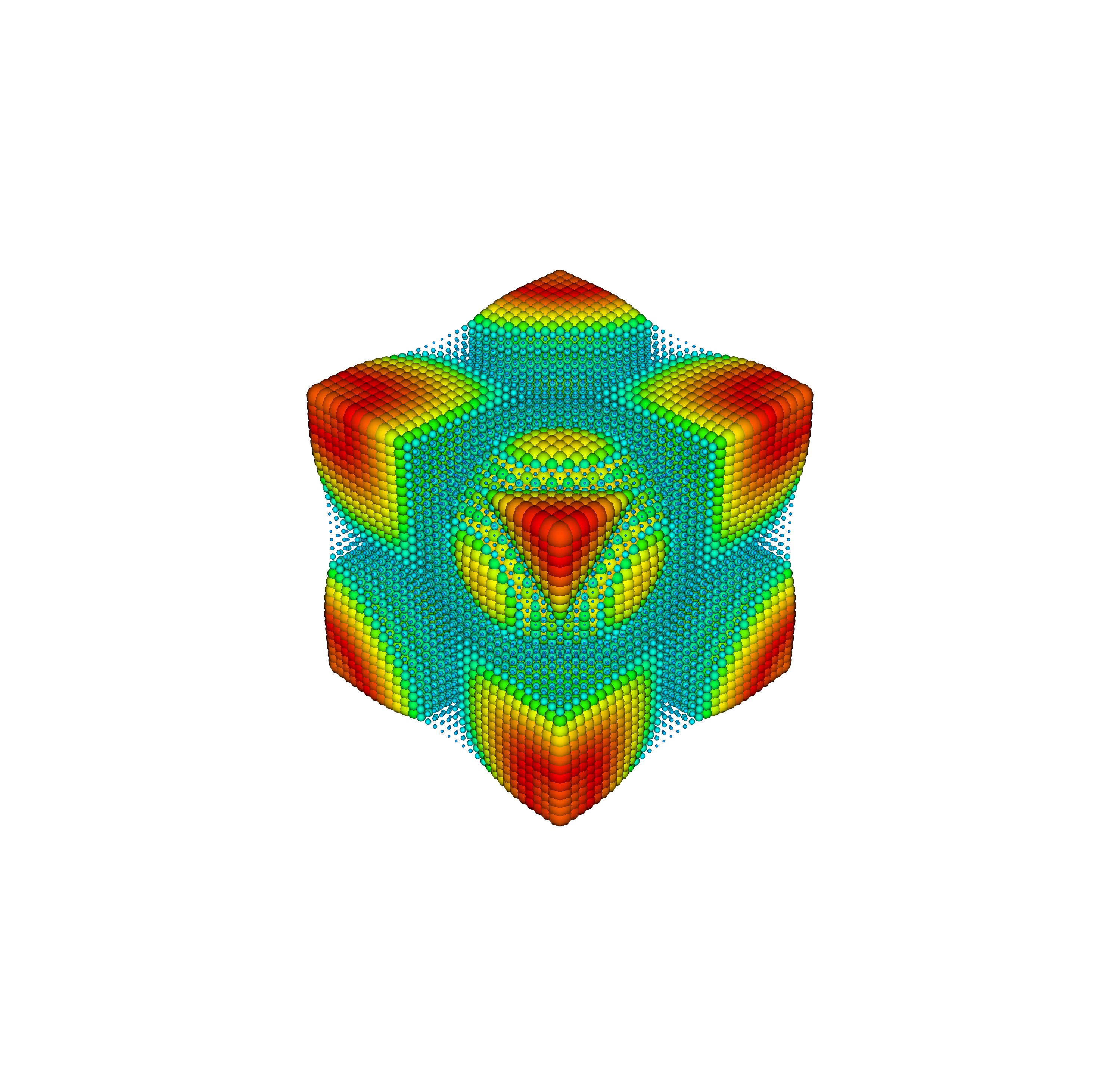}
          \caption{$\mathrm{SRCaIn_{2}}$ ($32\times 32\times 32$)}
          \vspace{2pt}
          \label{fig:SrCaIn2}
        \end{subfigure}%
        % \hfill
        \begin{subfigure}[b]{.3\linewidth}
          \centering
          \includegraphics[width=\linewidth]{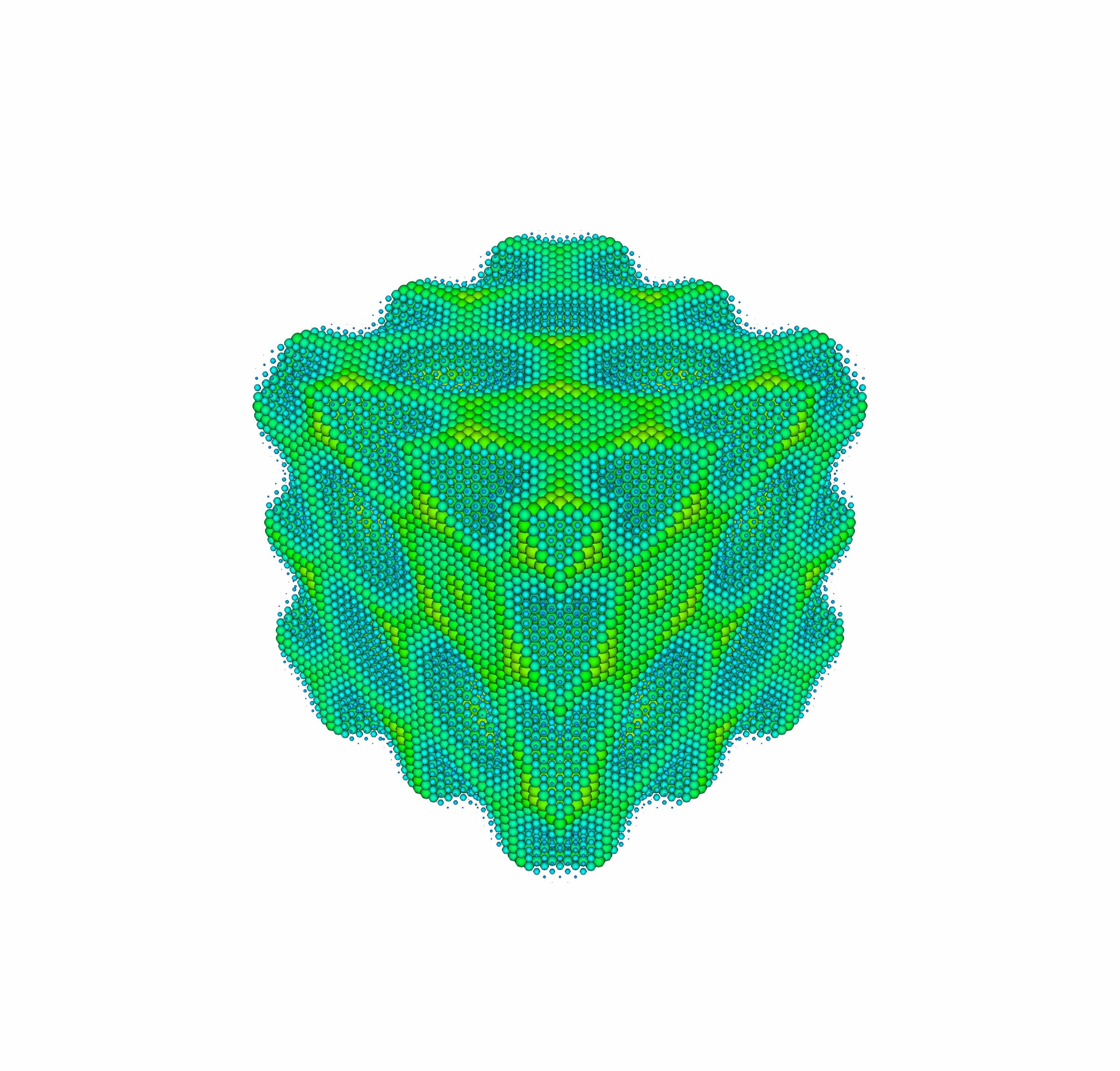}
          \caption{$\mathrm{Mn_{23}C_{6}}$ ($40\times 40\times 40$)}
          \vspace{2pt}
          \label{fig:Mn23C6}
        \end{subfigure}%
        % \hfill
        \begin{subfigure}[b]{.3\linewidth}
          \centering
          \includegraphics[width=\linewidth]{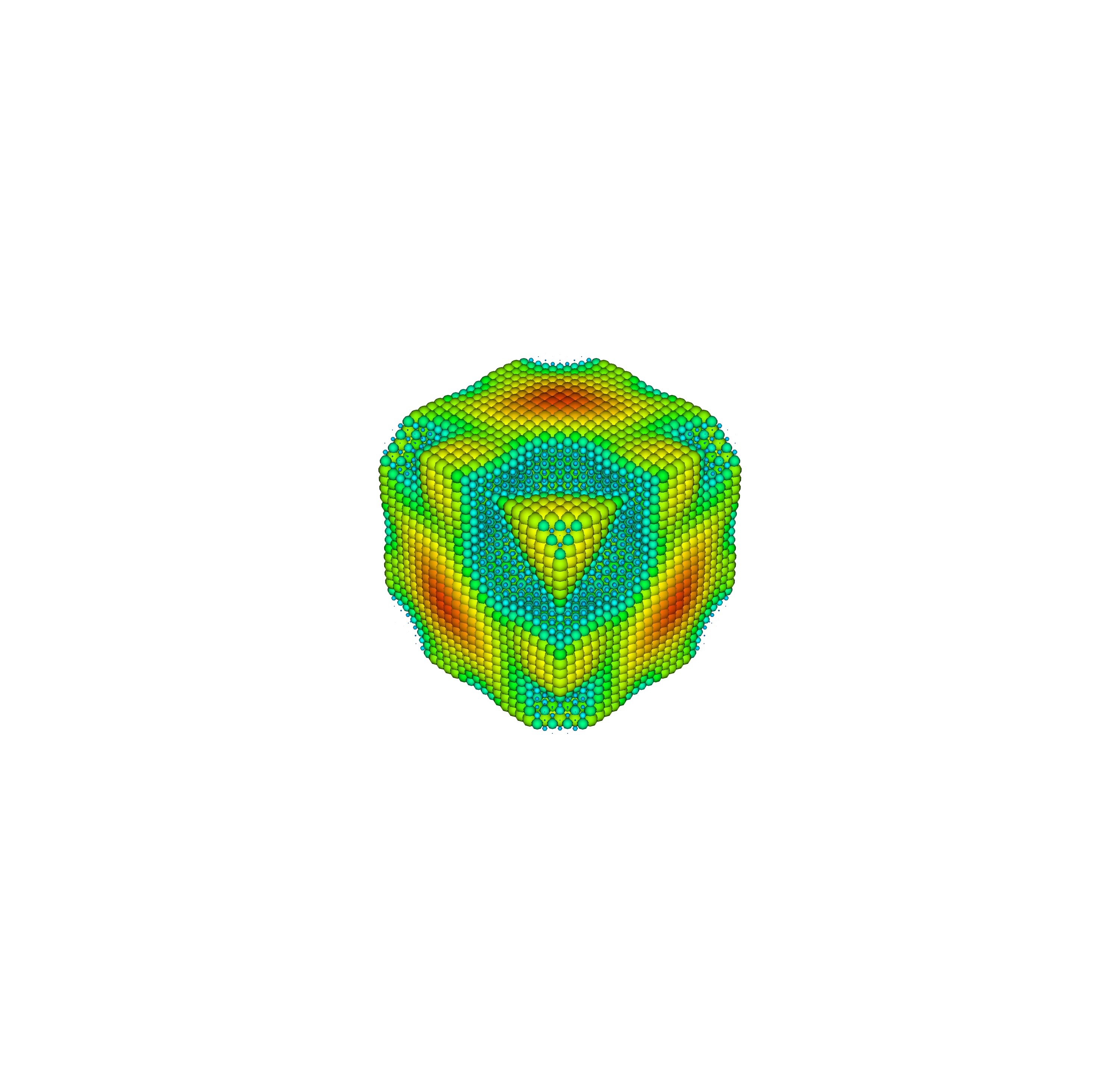}
          \caption{$\mathrm{VSiOs_{2}}$ ($24\times 24\times 24$)}
          \vspace{2pt}
          \label{fig:VSiOs2}
        \end{subfigure}
        % \hfill
        \begin{subfigure}[b]{.3\linewidth}
          \centering
          \includegraphics[width=\linewidth]{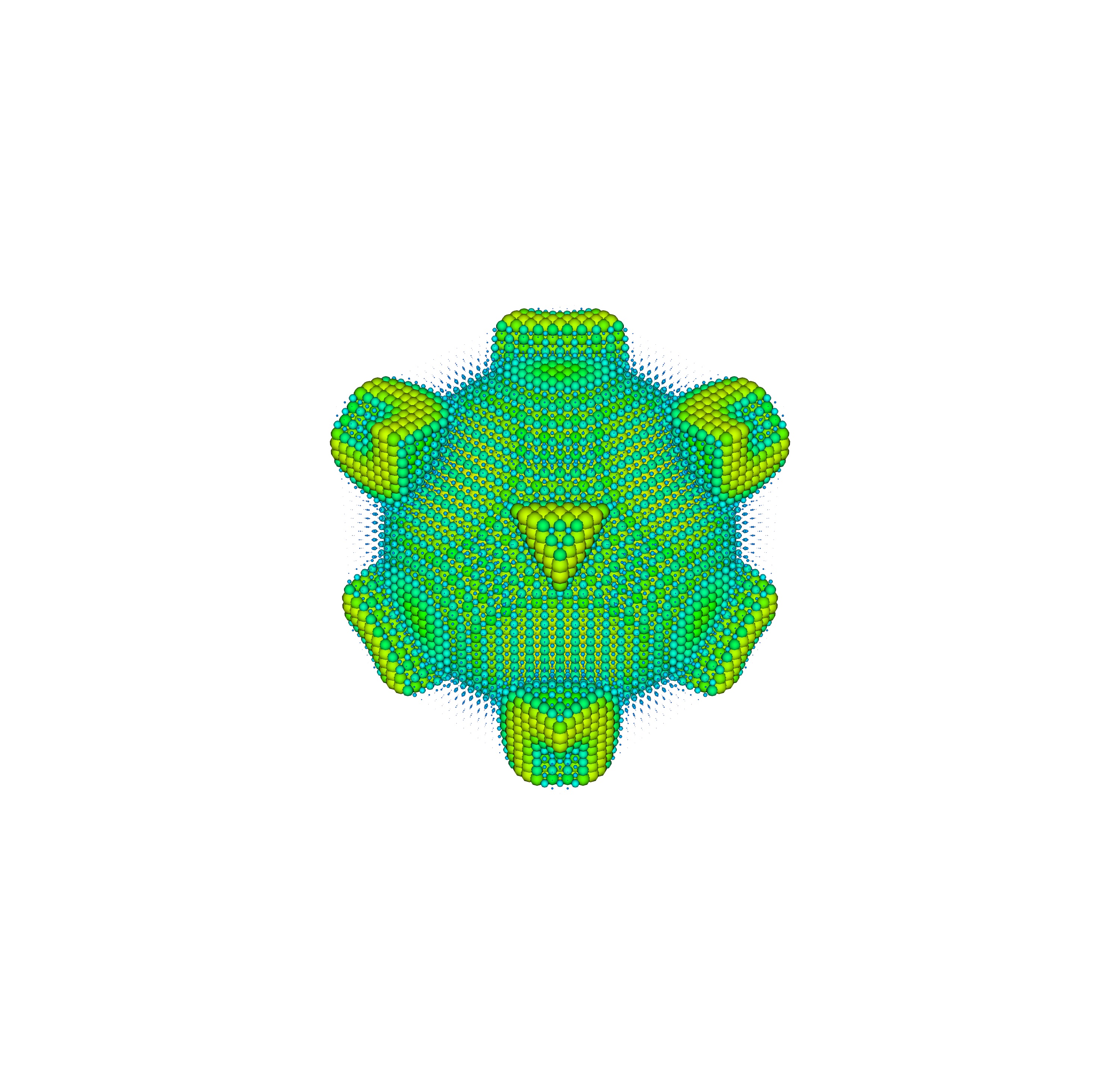}
          \caption{RbI ($30\times 30\times 30$)}
          \label{fig:RbI}
        \end{subfigure}%
        % \hfill
        \begin{subfigure}[b]{.3\linewidth}
          \centering
          \includegraphics[width=\linewidth]{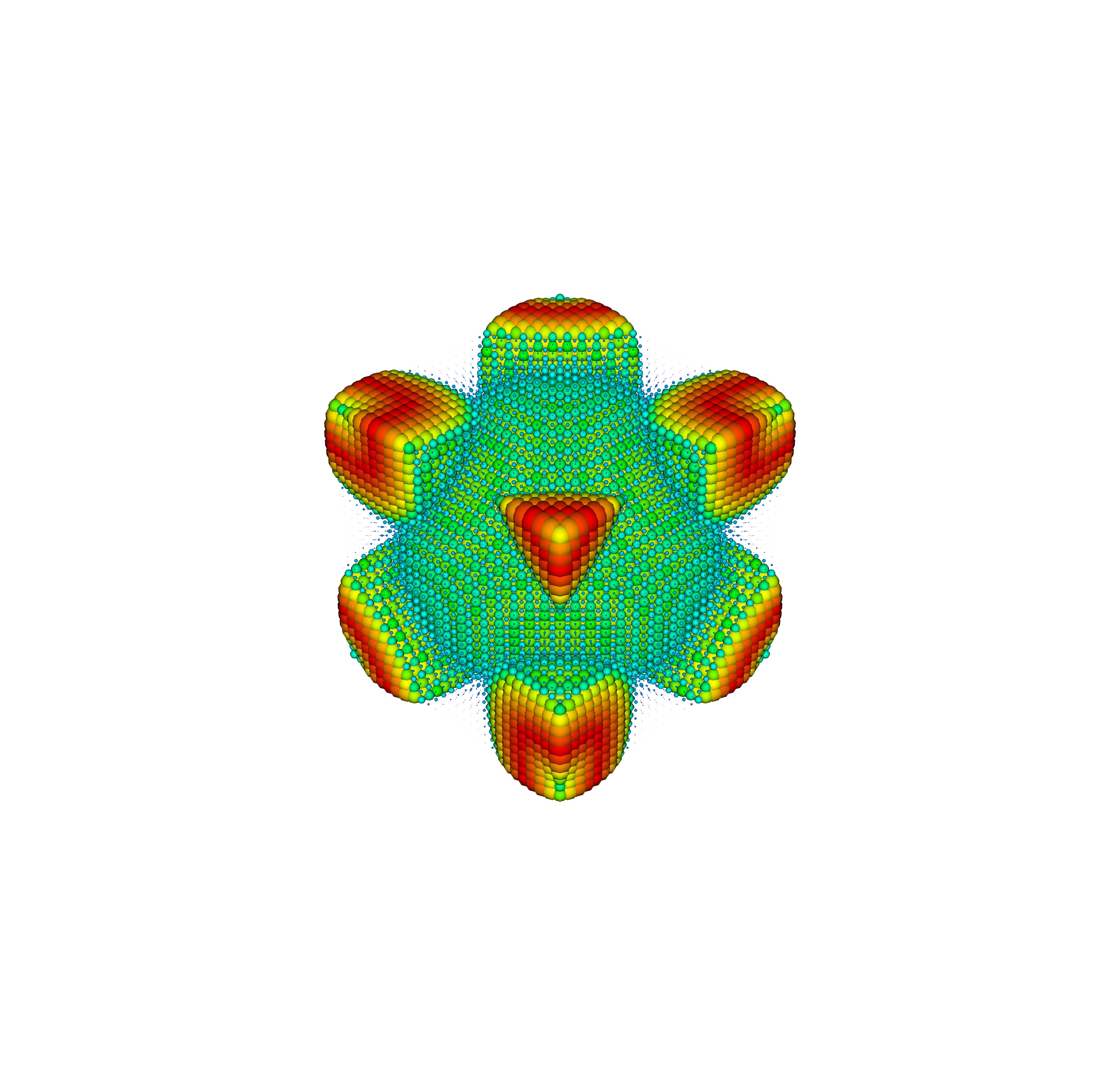}
          \caption{CsBr ($30\times 30\times 30$)}
          \label{fig:CsBr}
        \end{subfigure}%
        % \hfill
        \begin{subfigure}[b]{.3\linewidth}
          \centering
          \includegraphics[width=\linewidth]{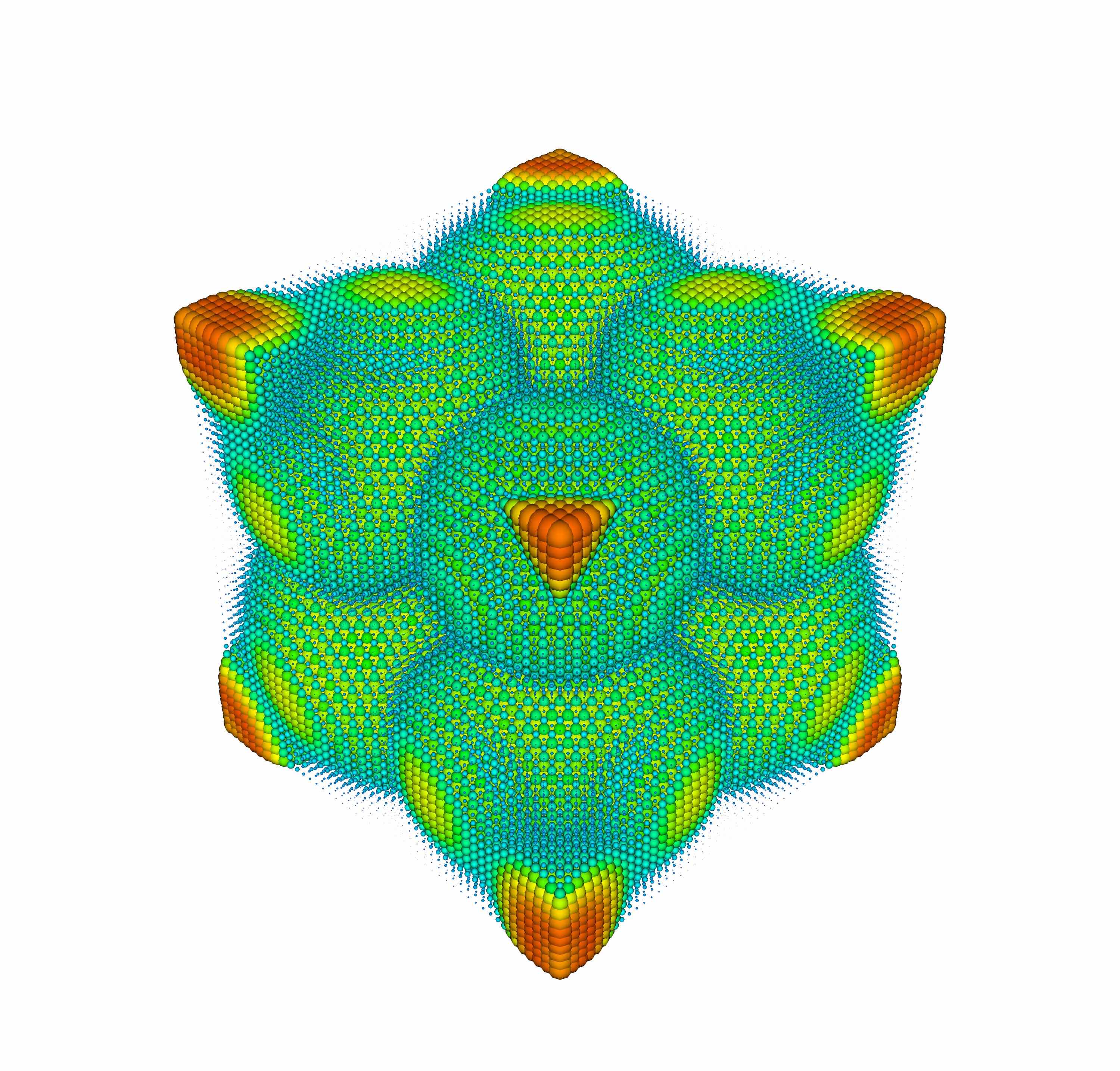}
          \caption{$\mathrm{VSiOs_{2}}$ ($48\times 48\times 48$)}
          \label{fig:Rb2TeBr6}
        \end{subfigure}
      \caption {Visualization of ECD for 6 materials showing clearly contrasting structural features (top and bottom rows). The top row are materials with high bulk modulus and the bottom row are materials with low bulk modulus. $l\times w\times h$ is the actual length, width and height of each ECD matrix.}
        \label{fig:visual_ECD}
    \end{figure}

\end{itemize}

\subsection{Machine Learning Methods}
In this work, we use Random Forest and Convolutional Neural Networks (CNNs) with Magpie features as the baseline methods. We propose that CNNs with ECD can capture certain characteristic relationships between material structures and their elastic properties.

Random Forest (RF)~\cite{breiman2001random} is a widely used machine learning model in material informatics because of its high accuracy and robustness~\cite{takahashi2019creating, oviedo2018fast, ward2017including}. As an ensemble learning algorithm, a RF aggregates the results from different decision trees (50 in this work). The decision trees are randomly trained based on subsets of training samples and features. Within a decision tree, a set of decision rules (e.g. Melting temperature > 200.0) is learned by minimizing the variance of the decision tree. For predicting elastic properties, RF calculates the final results by averaging outputs of all decision trees.

\begin{figure}[H]
  \centering
  \includegraphics[width=\linewidth]{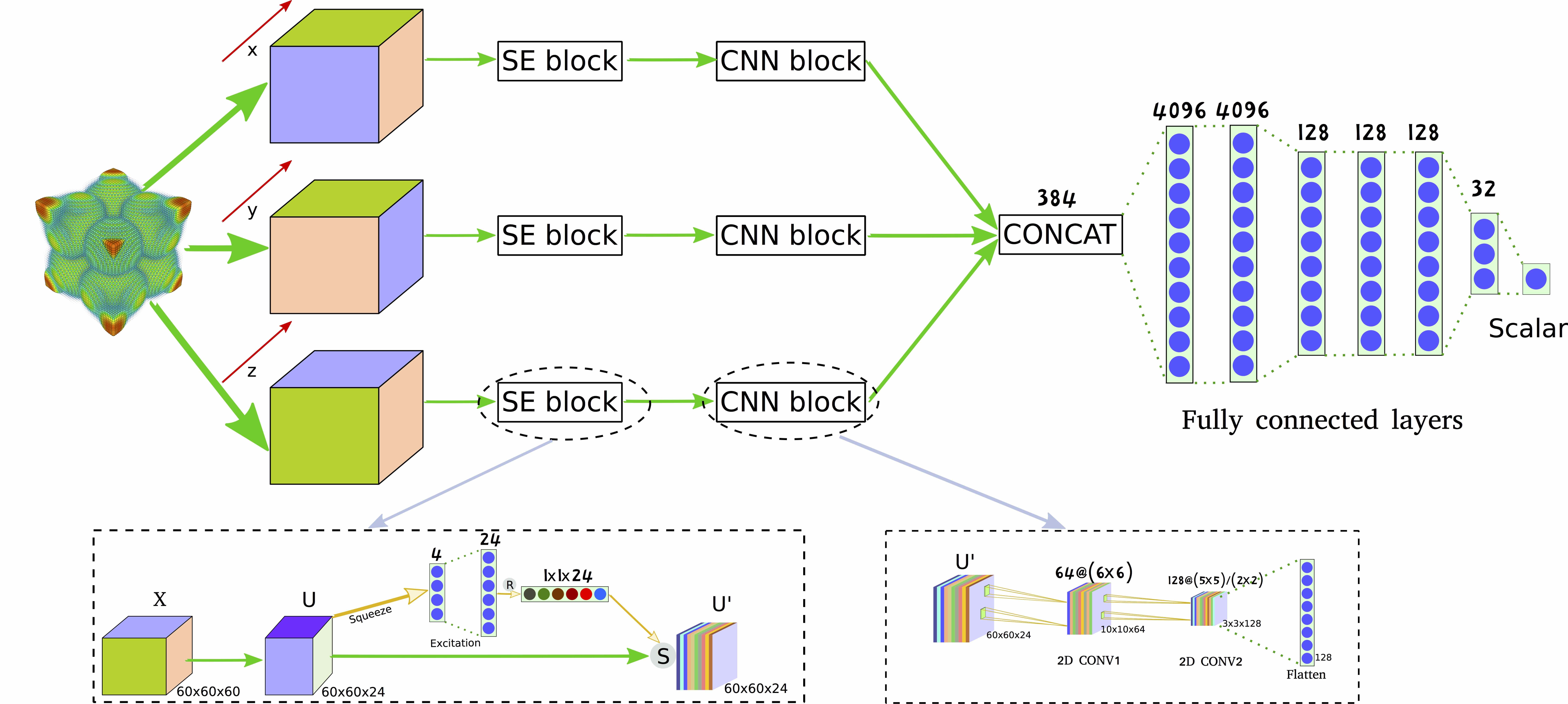}  
    \caption {The architecture of the 2D CNN with ECD representation. The "Scalar" stands for the bulk or shear modulus. The model includes three parts: mainframe, SE block and CNN block. In the mainframe, we have three branches whose outputs are concatenated and fed into six fully connected layers. The numbers above each component/layer are the number of neurons of that layer. In SE block, the labels of \textbf{R} and \textbf{S} are reshape and channel-wise multiplication operations. For simplicity, we ignore the max-pooling layers following every convolutional layer in the CNN block. Numbers below each component are the output dimension of that layer.}
    \label{fig:senet}
\end{figure}

Convolutional Neural Networks are a type of feed-forward neural network interleaved with convolutional, pooling, and fully connected layers. It has achieved state-of-the-art(SOTA) performance when applied to computer vision and natural language processing~\cite{simonyan2014very,krizhevsky2012imagenet,kim2014convolutional}. The convolutional unit is the core building block of CNNs, which is inspired by the multi-layered organization of the visual cortex. The unit consists of multiple learnable filters with a given receptive field and weight parameters. In our case, the filters are convolved across the full depth of the input volume of the ECD~\cite{lecun1998gradient}. The filters are learned hierarchically, where low-level features generate more condensed representations. The computational unit can be constructed by a transformation $U=F_{tr}(X), X\in \mathbb{R}^{L^{\prime}\times W^{\prime}\times H^{\prime}\times C^{\prime}}, U\in \mathbb{R}^{L\times W\times H\times C}$. $F_{tr}$ denotes the convolutional operation. Let $V=[v_{1}, v_{2},\dots, v_{C}]$ be the learnable convolutional filters. Then the outputs of $F_{tr}$ can be written as $U=[u_{1}, u_{2},\dots,u_{C}]$, where

\begin{equation} 
  u_{c}=v_{c}*X=\sum_{i=1}^{C^{\prime}}v_{c}^{i}*x^{i}
\end{equation}
Here $*$ denotes the dot product, $v_{c}=[v_{c}^{1},v_{c}^{2},\dots,v_{c}^{C^{\prime}}]$, $X=[x^{1},x^{2},\dots,x^{C^{\prime}}]$. We removed the bias terms for simplicity. $v_{c}^{i}$ is a 3D spatial filter convolving on a single channel of $X$. Stacked outputs of filters produce a 4D tensor activation map~\cite{lecun1998gradient}. A pooling layer is used to do non-linear downsampling. It partitions the 3D input into a set of rectangular boxes. In max-pooling, the pooling layer outputs the maximum value of each sub-region. Then a 3D tensor is activated through a rectified linear unit (ReLu)~\cite{krizhevsky2012imagenet}. The ReLu operation can be denoted by $max(0, P)$, where $P$ is the tensor generated by the max-pooling operation. The same procedure can be applied repeatedly to the whole activation map. Finally, the output of the convolutional layers is fed to one or more fully connected layers to accomplish the regression step. Similar procedures are applied in the CNN block in Figure~\ref{fig:senet}.

\begin{figure}[H]
  \centering
  \includegraphics[width=\linewidth]{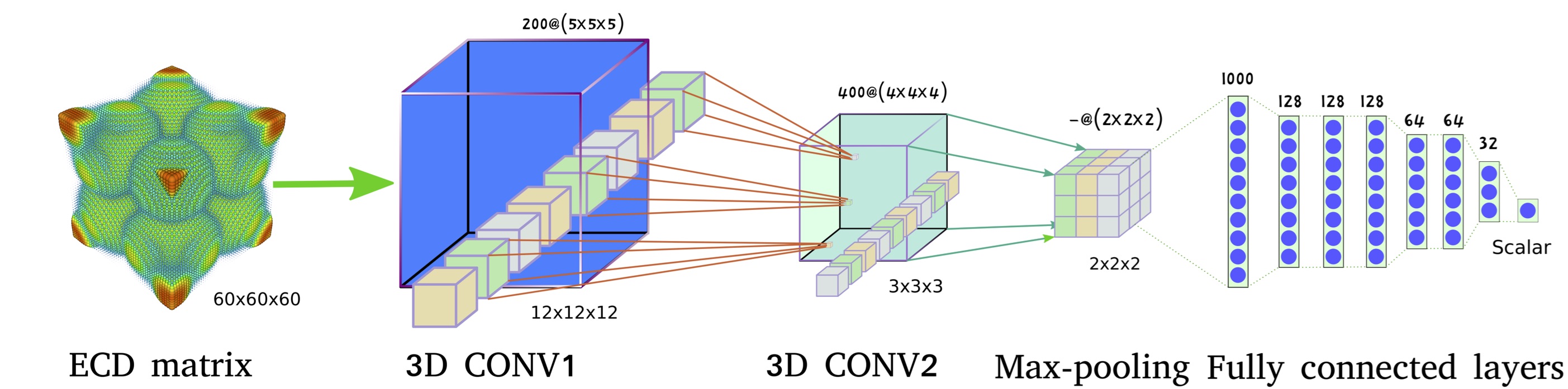}  
    \caption {The architecture of 3D CNN with ECD representations. The "Scalar" stands for the bulk or shear modulus. The numbers above each convolutional layers are its parameter settings. For instance, $200@(5\times 5\times 5)$ means 200 filters with size of $5\times 5\times 5$. Unless it is specified, the stride is always the same with the filter size. Two consecutive convolutional layers are followed by a max-pooling with pooling size and strides both of $2\times 2\times 2$. The number below are outputs of each layer. For fully connected layers, the numbers above them are the number of neurons.}
    \label{fig:3D_arch}
\end{figure}

We implemented two types of convolutional neural networks for learning ECD based features for elastic property prediction. Figure~\ref{fig:3D_arch} depicts the 3D CNN architecture in our work. This model has two consecutive convolutional layers followed by a max pooling layer, and then seven fully connected layers. For simplicity, we did not show the ReLu~\cite{krizhevsky2012imagenet} activation for all neural layers in Figure~\ref{fig:3D_arch}. The filter size of 2 convolution layers are $5\times 5\times 5$ and $4\times 4\times 4$, respectively and the stride has the same size as that of the convolution filters. For all max pooling layers, the sizes of filters and strides are $2\times 2\times 2$. The ECD matrices are fed to the 3D convolutional and pooling layers, and then the output matrix is flattened and passed to succeeding fully connected layers to calculate final predictions.

Figure~\ref{fig:senet} shows the architecture of our 2D CNNs for elastic property prediction. The ECD matrix does not have the concept of channel as images. Thus, we rotated the ECD matrix so that we can face the cube from x,y,z axes as shown in different colors of cubes in Figure~\ref{fig:senet}. Then the direction facing to us is considered as the channel direction. To model the inter-dependencies between channels, we used the Squeeze-and-Excitation (SE) network~\cite{hu2018squeeze}, which can exploit this inter-dependency by feature recalibration. This model selectively highlights the informative features and suppresses less useful ones. A SE block is shown in the left corner of Figure~\ref{fig:senet}. In this module, 24 filters of size of $1\times 1$ are used to down-sample the ECD matrix, which was first proposed in~\cite{lin2013network}. A nonlinearity operation is performed on each pixel across the channels. After the nonlinear projection, the ECD matrix $X$ of size $60\times 60\times 60$ is reduced to the feature map $U$ of size $60\times 60\times 24$.  A global average pooling is then used to shrink the feature map into a vector of size 24 along with the dimensions of width and height. Then we use a small set of fully connected layers to transform this vector into higher level features. The number of neurons on each layers are 4 and 24, respectively. The output $s$ of the last fully connected layer is reshaped into size of $1\times 1\times 24$. The last step is nonlinear excitation and the final output $U^{\prime}$ of block is achieved by rescaling the $U$ with the activated $s$:
  \begin{equation}
    U^{\prime} = U\odot \sigma(s)
  \end{equation}
where $\sigma$ is the sigmoid activation function that implements the nonlinear transformation. And $\odot$ denotes the channel-wise multiplication between the scalar $s$ and the feature map $U^{\prime}$.

The SE block in our 2D CNN architecture is followed by CNN blocks. A CNN block has two regular convolutional layers followed by a max-pooling layer. The first convolution neural has the same filter size and strides of $6\times 6$ and there are 64 filters in this layer. The second CNN layer has a filter size of $5\times 5$ and stride of $2\times 2$ and there are 128 filters in total. All max-pooling layers have the same pooling size and strides of $2\times 2$, respectively. 

For each of the projection map of x, y, and z, there is a SE and CNN block for feature extraction. The outputs of them are concatenated into a vector of size 384. Six fully connected layers are then used to map this learned features into elastic property values. The number of neurons on these fully connected layer are 4096, 4096, 128, 128, 128 and 32 respectively.

For the baseline algorithm, we also train a 2D CNN model with the Magpie features. To do that, we append 12 zeros to the Magpie features to get a vector of 1x144, which is then reshaped into a 2D matrix of size $12\times 12$. The CNN model for Magpie features has two consecutive convolutional layers followed by an average pooling layer. Then an additional convolutional layer is added followed by two fully connected layers. The  model parameters are set as follows: the kernel size and strides of the first convolutional layer are 3x3 and 1x1 and the number of filters is 32; the kernel size and strides of the second convolutional layer are 3x3 and 1x1 and the number of filters is 48; the pooling size and strides of the average pooling layer are both set as 2x2; the kernel size and strides of the third convolutional layer are 3x3 and 1x1 and the number of filters is 64; the number of neurons of the two fully connected layers are 48 and 32, respectively.

\subsection{Training and Implementation}

Figure~\ref{fig:3D_arch} shows the detailed architecture of our 3D CNN and its parameters. The models are implemented using the open-source libraries of TensorFlow (\url{https://www.tensorflow.org}) and Keras (\url{https://keras.io}). The performance of the models are evaluated using 5-fold cross validation. The input ECD has a shape of $60\times 60\times 60$ by interpolation for smaller matrices. The CNN for Magpie is also trained using the Adam optimizer~\cite{kingma2014adam} with a batch size of 32 and learning rate of 0.001. The 3D CNN model parameters are learned using the Adam optimizer~\cite{kingma2014adam} with a initial learning rate of 0.0005. For the 2D CNNs with ECD, we use the SGD optimizer to learn the model parameters. The initial learning rate is 0.001 and it drops by $0.5^{\floor{\frac{epoch}{10}}}$
, where $epoch$ is the current epoch. The mean absolute error (MAE) is used as the loss function for all three CNN models. The open source matminer (\url{https://hackingmaterials.lbl.gov/matminer/}) is utilized to calculate the Magpie features.

%%%%%%%%%%%%%%%%%%%%%%%%%%%%%%%%%%%%%%%%%%
\section{Results and Discussions}
In this section, we discuss the experiments demonstrating the potential of ECD for  material representation and elastic property prediction. The experiments are separated into two parts in terms of the evaluation approaches: experiments with 5-fold cross validation and experiments focusing on extrapolation performance evaluation. All experiments of CNN models are repeated 5 times and the result presented herein is the average of their outputs.

\subsection{5-fold cross validation experiments with redundant dataset}
Table~\ref{t:5fold} shows the results from 5-fold cross validation on the whole dataset with 2170 samples. We find that the baseline models using Magpie features are better than CNNs with ECD across all evaluation metrics for predicting bulk and shear moduli. Overall, RF with Magpie performs slightly better than CNNs with Magpie. Although $R^{2}$ of RF with Magpie is 0.001 lower than that of CNNs with Magpie in predicting bulk modulus, RF with Magpie achieves much better results in predicting shear modulus ($R^{2}$ is 0.049 higher). Similar observations apply to performance evaluated in terms of Root Mean Square Error (RMSE). This better performance of Magpie based RF models are not unexpected. First, all samples in this FCC2170 dataset belong to the Fm$\bar{3}$m space group. By sharing similar structures, the Magpie features are able to capture most of the elastic property variation due to composition difference. The high structural similarity of the dataset helps the baseline methods based on composition Magpie features predict the elastic properties well. Another reason is that the FCC2170 contains many similar samples in terms of compositions. The high redundant samples also makes the baseline models with Magpie features to make precise predictions by exploiting redundant neighbor samples in the training set when evaluated on the test set during cross-validation. However, the machine learning models trained with redundant training set can lead to low extrapolation performance as shown in our previous study \cite{xiong2020evaluating}.

Here we show that ECD can be used as a complementary materials descriptor for elastic property prediction together with the Magpie features. To verify this, We pre-trained a CNN model with Magpie features and a 2D CNN model with ECD. Then we fused these two models by concatenating the outputs of the penultimate layers of these two models to generate a output latent feature vector of dimension 64, which is then fed to three fully connected layers with 128, 64, and 32 neurons respectively. The Adam optimizer~\cite{kingma2014adam} is used for training with a learning rate of 0.001. This fusion neural network model with mixed Magpie and ECD descriptor yielded the best $R^{2}$ and RMSE of 0.955 (0.804) and 16.530 (15.780) in predicting bulk (shear) modulus respectively as shown in Table~\ref{t:5fold} . This confirms that ECD and Magpie can work together to achieve better performance for elastic property prediction. In addition, our experiments also showed that the projected 2D CNN achieved significantly  better performance than the basic 3D CNN models. The $R^{2}$ and RMSE of 2D-CNN with ECD are 0.912 and 23.401 in predicting bulk modulus compared to 0.884 and 26.819 of 3D-CNNs with ECD. The $R^{2}$ and RMSE of 2D CNN with ECD are 0.768 and 17.192 in predicting shear modulus compared to 0.745 and 17.944 of 3D-CNNs with ECD.

\begin{table}[H]
  \centering
  \caption{Performance Comparisons of models with Magpie and ECD descriptors using 5-fold cross validation}
  \begin{tabular}{|c|c|c|c|c|c|c|c|c|c|c|}
  \hline
  \multicolumn{1}{|c|}{\multirow{2}{*}{Type}} & \multicolumn{2}{c|}{RF+Magpie} & \multicolumn{2}{c|}{CNN+Magpie} & \multicolumn{2}{c|}{3D-CNN+ECD}& \multicolumn{2}{c|}{2D CNN+ECD}& \multicolumn{2}{c|}{Fusion} \\ \cline{2-11} 
  \multicolumn{1}{|c|}{}&$R^{2}$&RMSE &$R^{2}$&RMSE &$R^{2}$&RMSE &$R^{2}$&RMSE &$R^{2}$&RMSE\\ \hline

  bulk&0.943&18.721&0.944&18.423&0.884&26.819&0.912&23.401&\textbf{0.955}&\textbf{16.530}\\ \hline

  shear&0.794&16.142&0.745&17.959&0.745&17.944&0.768&17.192&\textbf{0.804}&\textbf{15.780}\\ \hline
  \end{tabular}
  \label{t:5fold}
\end{table}

\subsection{Extrapolation Experiments with non-redundant datasets}
ML models with elemental descriptors such as Magpie can achieve good cross-validation performance for datasets consisting of redundant (computationally very similar samples) such as FCC2170. However, the better performance of the fusion model with CNN with Magpie and 2D-CNN with ECD implies that for the ECD descriptor can help to make better predictions over a certain subset of test samples. In this section, we aim to construct non-redundant dataset and show that our CNN models with the ECD descriptor can achieve better performance on non-redundant datasets or for test samples with few highly similar neighbor samples. 

For these extrapolation experiments, we trained and tested the prediction models over the FCC-E-N datasets as described in Section~\ref{section:dataset}. The performance comparison results of the extrapolation experiments for bulk and shear modulus prediction are shown in Table~\ref{t:minus_elem}.  There are 22 sets of experiments with 11 of them for predicting bulk modulus and the other 11 for predicting shear modulus by five different algorithms including RF+Magpie, CNN+Magpie, 3D-CNN+ECD, 2D-CNN+ECD, and the latest crystal graph convolutional neural network (CGCNN) \cite{xie2018crystal}, which also uses structural information. We highlighted the best performance scores for each experiments and count how many experiments each algorithm achieved the best scores. As shown in Table~\ref{t:minus_elem}, the RF with Magpie and CNN with Magpie worked the best for 5 and 6 experiments respectively. However, impressively, for these non-redundant training/testing experiments, our ECD descriptor based 3D-CNN-ECD and 2D-CNN-ECD outperformed the others for 4 and 5 experiments respectively, which reflecting the importance of the structure based ECD descriptor for elastic property prediction. In contrast, the popular CGCNN only achieved the best performance out of 2 experiments, which demonstrated the advantage of our ECD based atomic structure representation. 

\begin{table}[H]
  \centering
  \caption{Extrapolation prediction performance comparison on non-redundant leave-one-element-out datasets}
  \begin{tabular}{|c|c|c|c|c|c|c|c|c|c|c|c|}
  \hline
  \multirow{2}{*}{Elem} & \multirow{2}{*}{Type} & \multicolumn{2}{c|}{RF+Magpie}& \multicolumn{2}{c|}{CNN+Magpie}  & \multicolumn{2}{c|}{3D-CNN+ECD} & \multicolumn{2}{c|}{2D-CNN+ECD} & \multicolumn{2}{c|}{CGCNN}\\ \cline{3-12} 
    & &$R^{2}$  & RMSE & $R^{2}$ & RMSE & $R^{2}$ & RMSE  & $R^{2}$ &RMSE& $R^{2}$ &RMSE\\ \hline

  \multirow{2}{*}{F} &bulk& -0.529&26.797&-0.809&29.102&\textbf{-0.051}&\textbf{22.212}&-0.448&26.080&-2.217&35.554\\ \cline{2-2} \cline{3-12} 
                    &shear&-3.350&18.117&-6.912&24.315&-1.202&12.878&-1.293&13.151&\textbf{-0.548}&\textbf{10.657}\\ \hline

  \multirow{2}{*}{K} &bulk&\textbf{0.776}&\textbf{6.067}&0.646&7.573&0.510&8.969&0.570&8.397&0.474&9.055\\ \cline{2-2} \cline{3-12} 
                    &shear&\textbf{0.810}&\textbf{2.641}&0.548&4.014&0.389&4.733&0.367&4.817&0.146&5.523\\ \hline

  \multirow{2}{*}{Rb} &bulk&0.867 &4.603&\textbf{0.869}&\textbf{4.579}&0.753&6.287& 0.777& 5.966&0.275&10.290\\ \cline{2-2} \cline{3-12} 
                    &shear&\textbf{0.778}&\textbf{2.767}&0.727&3.064&0.608&3.657&0.719&3.111 &0.268&4.944 \\ \hline

  \multirow{2}{*}{Cs} &bulk& -0.128& 11.232&\textbf{0.760}&\textbf{5.166}&0.448&7.818&0.067&10.158&-0.144&10.934\\ \cline{2-2} \cline{3-12} 
                    &shear&-4.327&11.199&\textbf{0.492}&\textbf{3.446}&0.014&4.743&-1.137&7.083&0.344&3.881\\ \hline

  \multirow{2}{*}{Na} &bulk& 0.630&16.398&\textbf{0.833}&\textbf{11.013}&0.660&15.708&0.616&16.689&0.605&16.223\\ \cline{2-2} \cline{3-12} 
    &shear& 0.545&8.366&0.386&9.716&\textbf{0.548}&\textbf{8.340}&0.451&9.196&0.351&9.863\\ \hline 

  \multirow{2}{*}{Cl} &bulk& 0.410&15.935&0.529&14.151&0.591&13.009&\textbf{0.716}&\textbf{11.05}&0.534&13.119\\ \cline{2-2} \cline{3-12} 
                    &shear&-0.477&10.715&0.213&7.765&\textbf{0.339}&\textbf{7.160}&0.093&8.394&-0.197&9.366\\ \hline

  \multirow{2}{*}{In} &bulk& \textbf{0.829}&\textbf{20.780}&0.780&23.550&0.725&26.326&0.773&23.908&0.761&24.460\\ \cline{2-2} \cline{3-12} 
                    &shear&0.791&8.250&0.771&8.618&0.683&10.136&\textbf{0.793}&\textbf{8.207}&0.655&10.416\\ \hline

  \multirow{2}{*}{Br} &bulk &0.921&4.464&0.923&4.585&0.912&4.700&\textbf{0.923}&\textbf{4.411}&0.631&9.245\\ \cline{2-2} \cline{3-12} 
                    &shear& 0.630 &2.290&-0.078&3.857&0.755&1.861&\textbf{0.824}& \textbf{1.579}&-2.661&6.975\\ \hline

  \multirow{2}{*}{Li}&bulk  &0.418&29.869&\textbf{0.867}&\textbf{14.253}&0.519&27.142&0.454&28.937&0.732&20.121\\ \cline{2-2} \cline{3-12} 
                    &shear&-0.232&17.799&0.416&12.239&0.428&12.126&\textbf{0.451}&\textbf{11.881}&0.388&12.488\\ \hline

  \multirow{2}{*}{Sc} &bulk&0.855&23.276&\textbf{0.908}&\textbf{18.538}&0.756&30.195& 0.850& 23.688&0.818&25.983\\ \cline{2-2} \cline{3-12} 
                    &shear&\textbf{0.781}&\textbf{12.996}&0.707&15.024&0.682&15.650&0.635&16.786&0.667&16.007\\ \hline

  \multirow{2}{*}{Tl} &bulk& -0.370&24.574&0.421&15.973&0.219&18.529&0.501&14.833&\textbf{0.550}&\textbf{14.040}\\ \cline{2-2} \cline{3-12} 
                    &shear&0.456& 6.745&0.437&6.815&\textbf{0.559}&\textbf{6.068}&0.557&6.084&0.427&6.818\\ \hline

  \multicolumn{2}{|c|}{\textbf{\# of the best}} & \multicolumn{2}{c|}{\textbf{5}} & \multicolumn{2}{c|}{\textbf{6}} & \multicolumn{2}{c|}{\textbf{4}} & \multicolumn{2}{c|}{\textbf{5}}& \multicolumn{2}{c|}{2}  \\ \hline
  \end{tabular}
  \label{t:minus_elem}
\end{table}

\subsection{Visualization Study of when ECD descriptor works better}
To understand on why our ECD based CNN models worked better than Magpie features on some datasets but not others, we conducted a visualization study for all the extrapolation experiments.
For magpie features, we directly apply the t-distributed Stochastic Neighbor Embedding (t-SNE)~\cite{maaten2008visualizing} to the dataset. For the ECD based features, directly applying t-SNE is not feasible due to the memory limit. So we first applied max-pooling to the 3D ECD matrices with strides of $(6, 6, 6)$ and pooling size of $(6, 6, 6)$ before feeding them into t-SNE. Hence the final size of the ECD matrices is $(10, 10, 10)$, which are then flattened to a 1D vector of 1,000 elements. Then we applied t-SNE to this 1D vector to reduce the dimension to 2.

\begin{figure}[H]
    \centering
    \begin{subfigure}[b]{0.475\textwidth}   
        \centering 
        \includegraphics[width=\textwidth]{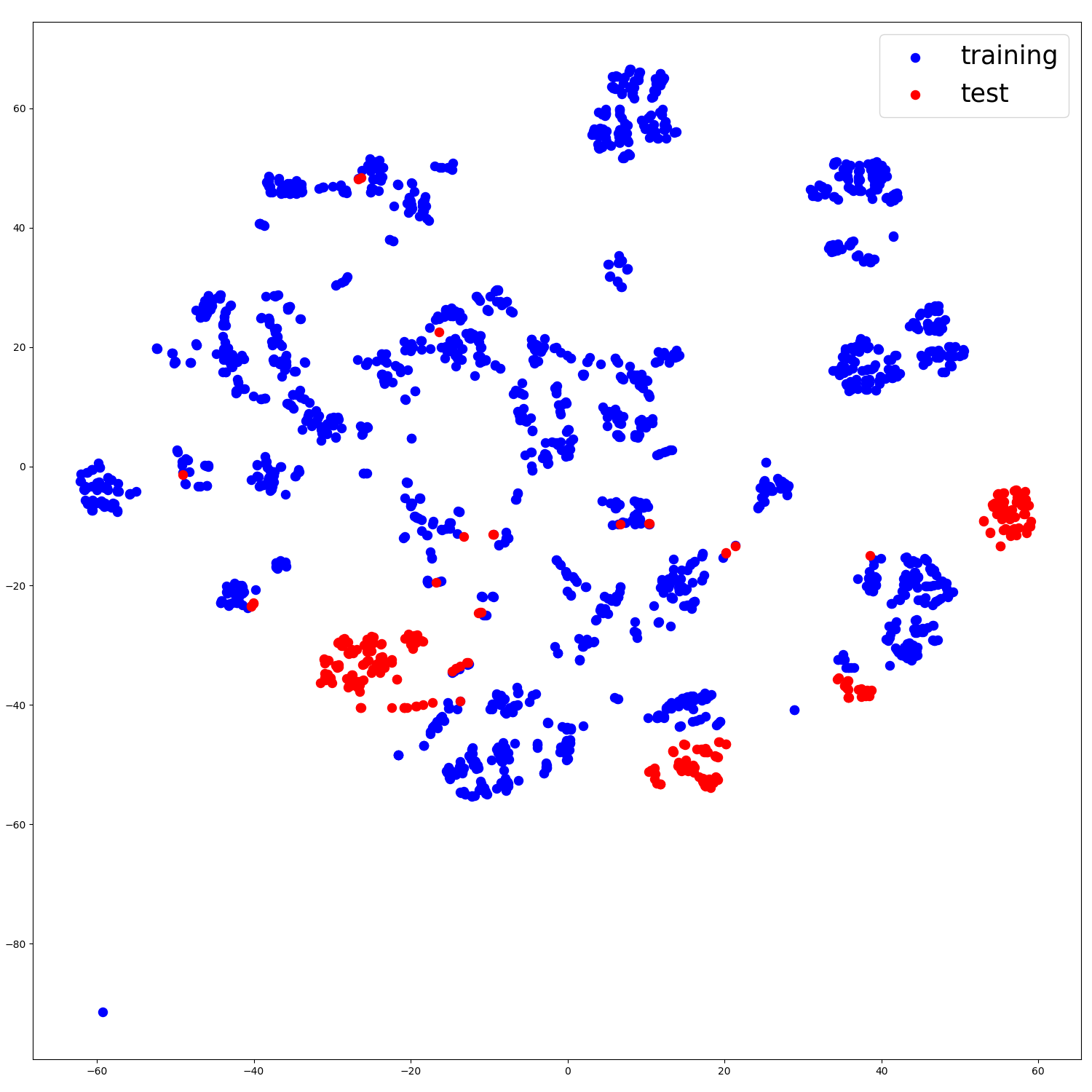}
        \caption[]%
        {{\small 2D map of Magpie features for FCC-Cl-1880}}    
        \label{fig:magpie_Cl}
    \end{subfigure}
    % \quad
    \begin{subfigure}[b]{0.475\textwidth}   
        \centering 
        \includegraphics[width=\textwidth]{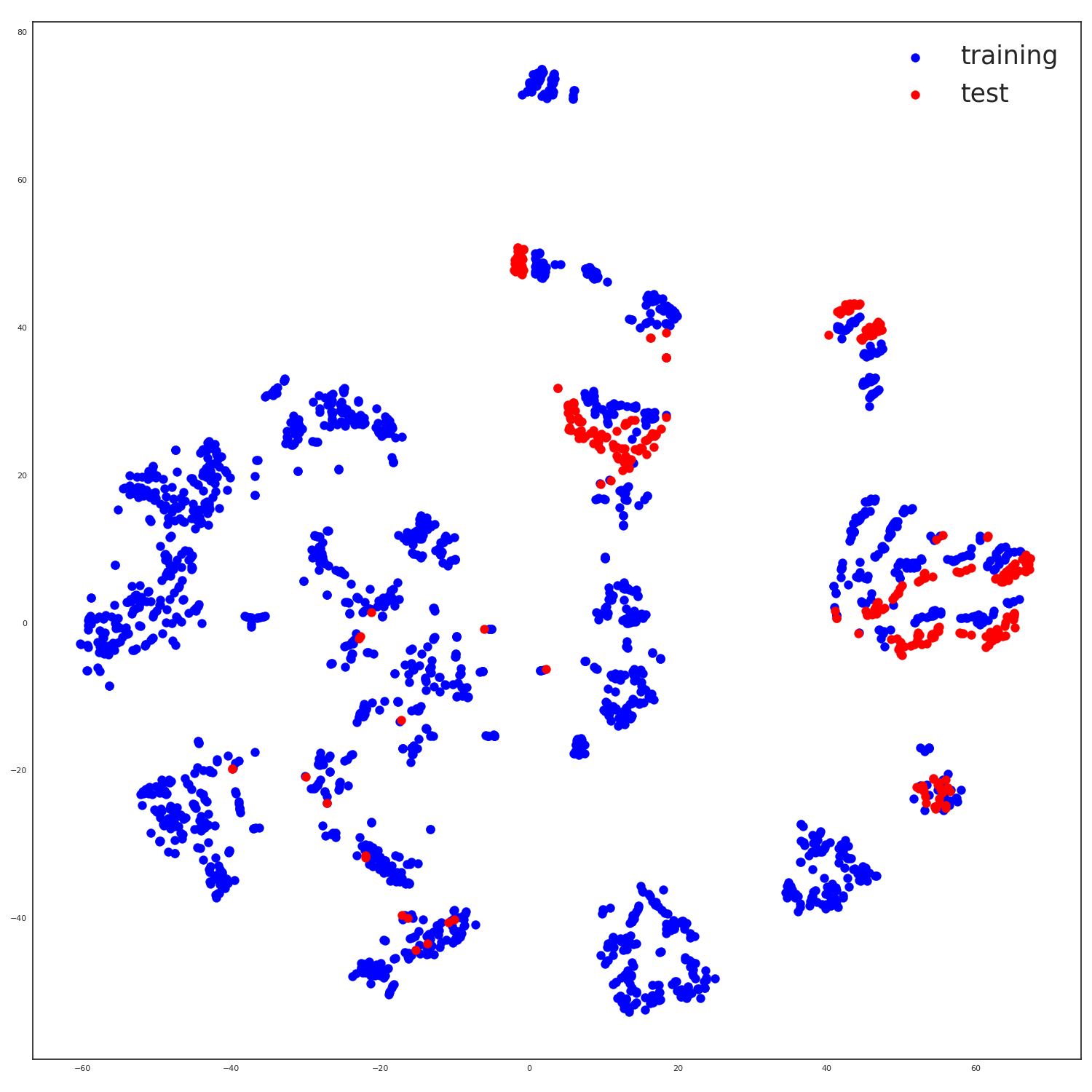}
        \caption[]%
        {{\small 2D map of ECD features for FCC-Cl-1880}}    
        \label{fig:ecd_Cl}
    \end{subfigure}

    \vskip\baselineskip
    
    \begin{subfigure}[b]{0.475\textwidth}
        \centering
        \includegraphics[width=\textwidth]{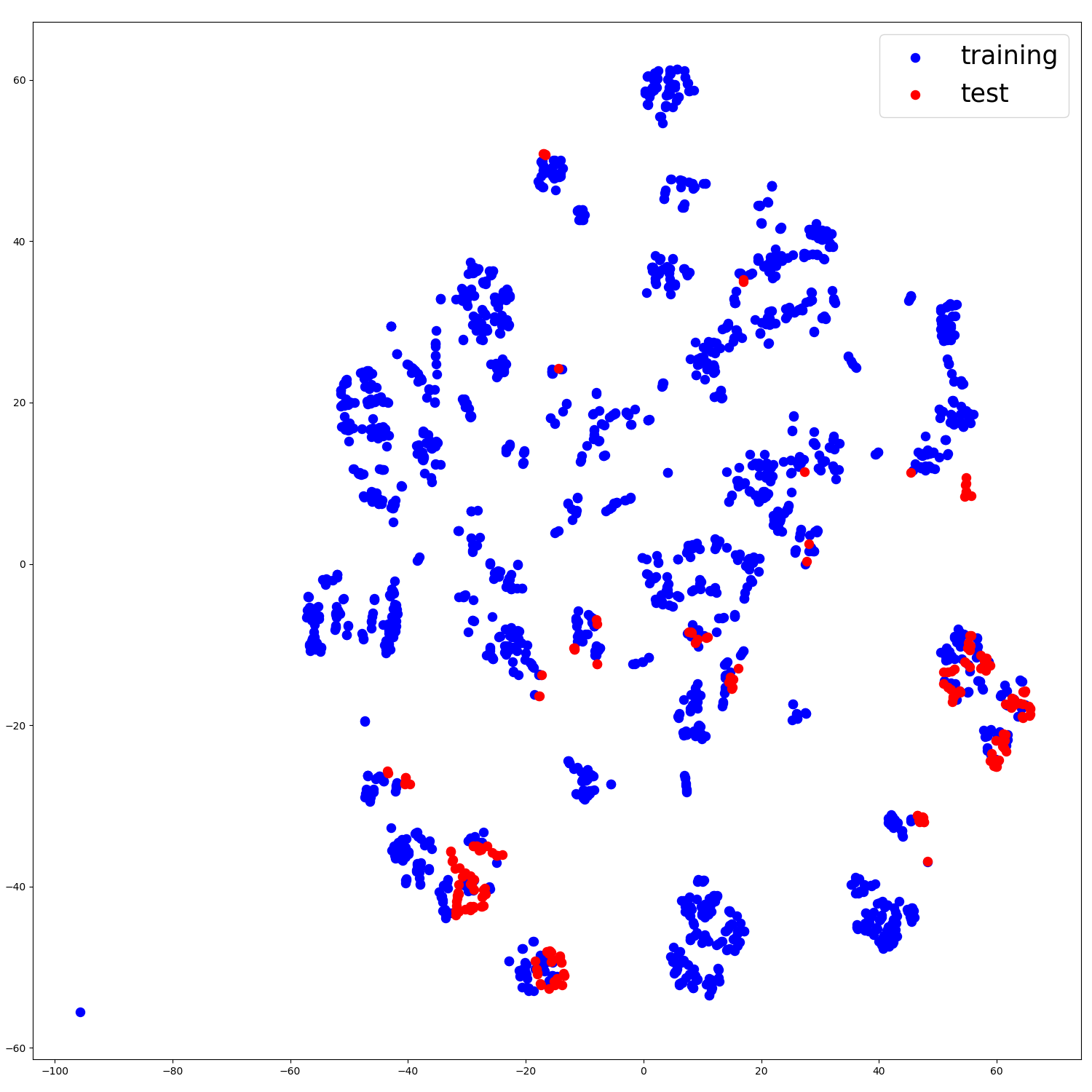}
        \caption[]%
        {{\small 2D map of Magpie features for FCC-Tl-1966}}    
        \label{fig:magpie_Tl}
    \end{subfigure}
    % \hfill
    \begin{subfigure}[b]{0.475\textwidth}  
        \centering 
        \includegraphics[width=\textwidth]{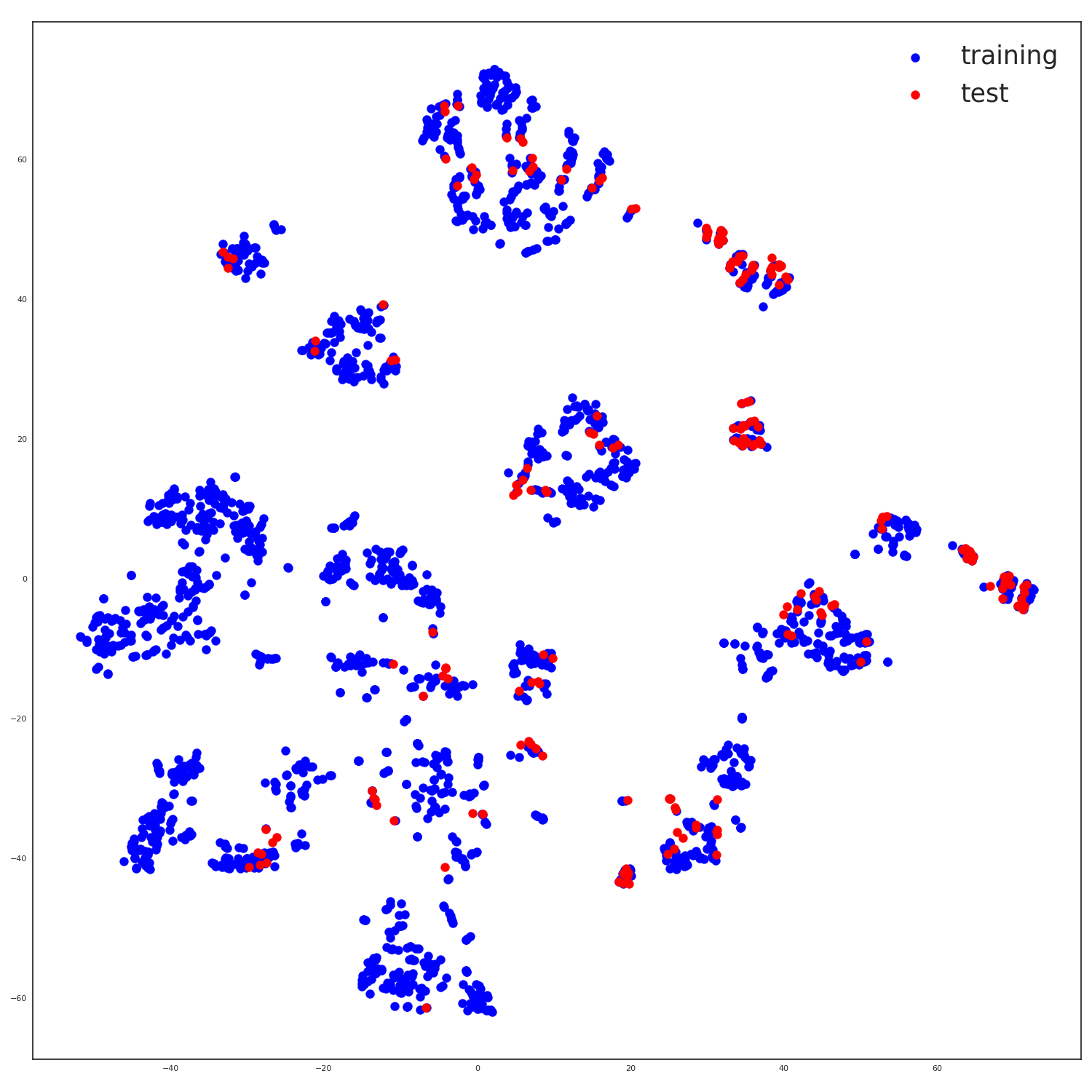}
        \caption[]%
        {{\small 2D map of ECD features for FCC-Tl-1966}}    
        \label{fig:ecd_Tl}
    \end{subfigure}
    \caption[]
    {\small Visualization of high-dimensional features for element Chlorine and Thallium by t-SNE. Blue dots are training data and red dots are test data. } 
    \label{fig:group_elem}
\end{figure}

Figure~\ref{fig:group_elem} shows 2D visualization of the high-dimension Magpie and ECD features for two datasets: FCC-Chlorine-1880 and FCC-Thallium-1966 over which the ECD based models outperform Magpie feature based models. The training samples are labelled as blue points while the test samples are red points. First, Figure~\ref{fig:group_elem} (a) and (b) show the distribution of training and test samples with Magpie features and with ECD features respectively for the FCC-Chlorine-1880 dataset. In subfigure~\ref{fig:magpie_Cl}, we found that there exist three large clusters of test samples (red points) that have few similar training samples around. This corresponds to the low prediction performance for Magpie based models. The best performance for both bulk and shear modulus prediction is achieved by CNN+Magpie with R2 of 0.529 and 0.213 respectively. In contrast, subfigure~\ref{fig:ecd_Cl} shows the 2D distribution of the samples represented with ECD features. It can be found that the test samples are mostly mixed with training samples, leading to much better prediction performance: the best performance for bulk modulus prediction is achieved by 2D-CNN+ECD with R2 of 0.716, which is significantly better (35\%)than 0.529, the best prediction performance achieved by Magpie based models. The best performance for shear prediction is achieved by 3D-CNN+ECD with R2 of 0.339, which is also 59\% better than 0.213, the best R2 score of Magpie based models.  

Figure~\ref{fig:group_elem} (c) and (d) show the distribution of training and test samples with Magpie features and with ECD features respectively for the FCC-Thallium-1966 dataset. In subfigure~\ref{fig:magpie_Tl}, we found that clusters of test samples (red points) are closer to training samples compared to subfigure~\ref{fig:magpie_Cl}. There is no large clusters of isolated test samples. The best performance for bulk modulus is achieved by CNN+Magpie with R2 of 0.421. The best performance for shear modulus prediction is achieved by RF+Magpie with R2 of 0.456. In contrast, subfigure~\ref{fig:ecd_Tl} shows the 2D distribution of the samples represented with ECD features. It can be found that the test samples are better mixed with training samples than subfigure~\ref{fig:magpie_Cl}, leading to better prediction performance. The best performance for both bulk modulus prediction is achieved by 2D-CNN+ECD with R2 of 0.501 and the best shear modulus prediction performance is achieved by 3D-CNN+ECD with R2 of 0.559. In this dataset, the best ECD based model is (0.559-0.421)/0.421 = 19\% better than the best Magpie based model for bulk modulus prediction. The performance gap is much smaller compared to that (35\%) on the FCC-Chlorine-1880 dataset. The best ECD based model is also (0.559-0.456)/0.456 = 24.9\% better than the best Magpie based model for shear modulus prediction, which is however much smaller than the performance gap over the FCC-Chlorine-1880 dataset, which is 59\%. These findings can partially explain why ECD based models are superior to Magpie based models in predicting elastic properties for these two datasets. It shows the structure based ECD descriptor can be a complementary descriptor to elemental Magpie features for elastic property prediction due to their better neighborhood structure of the samples. This analysis is consistent to those observation that neighbor sample distribution significantly affects the performance of neural network based prediction models \cite{jon2019}.

\subsection{DFT validation}
\begin{figure}[H]
    \centering
    \begin{subfigure}[b]{0.475\textwidth}   
        \centering 
        \includegraphics[width=\textwidth]{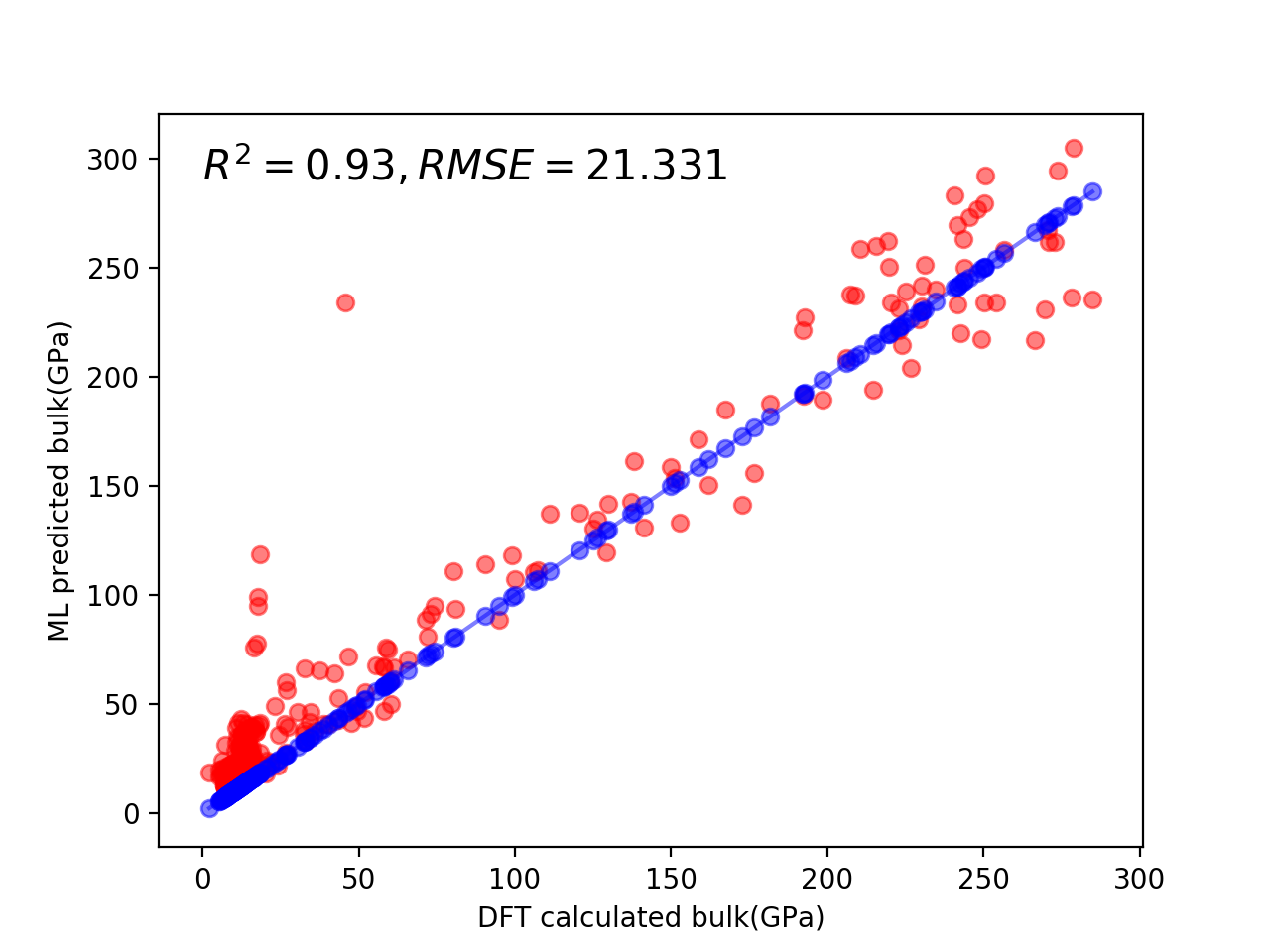}
        \caption[]%
        {{}}    
        \label{fig:bulk_parity}
    \end{subfigure}
    % \quad
    \begin{subfigure}[b]{0.475\textwidth}   
        \centering 
        \includegraphics[width=\textwidth]{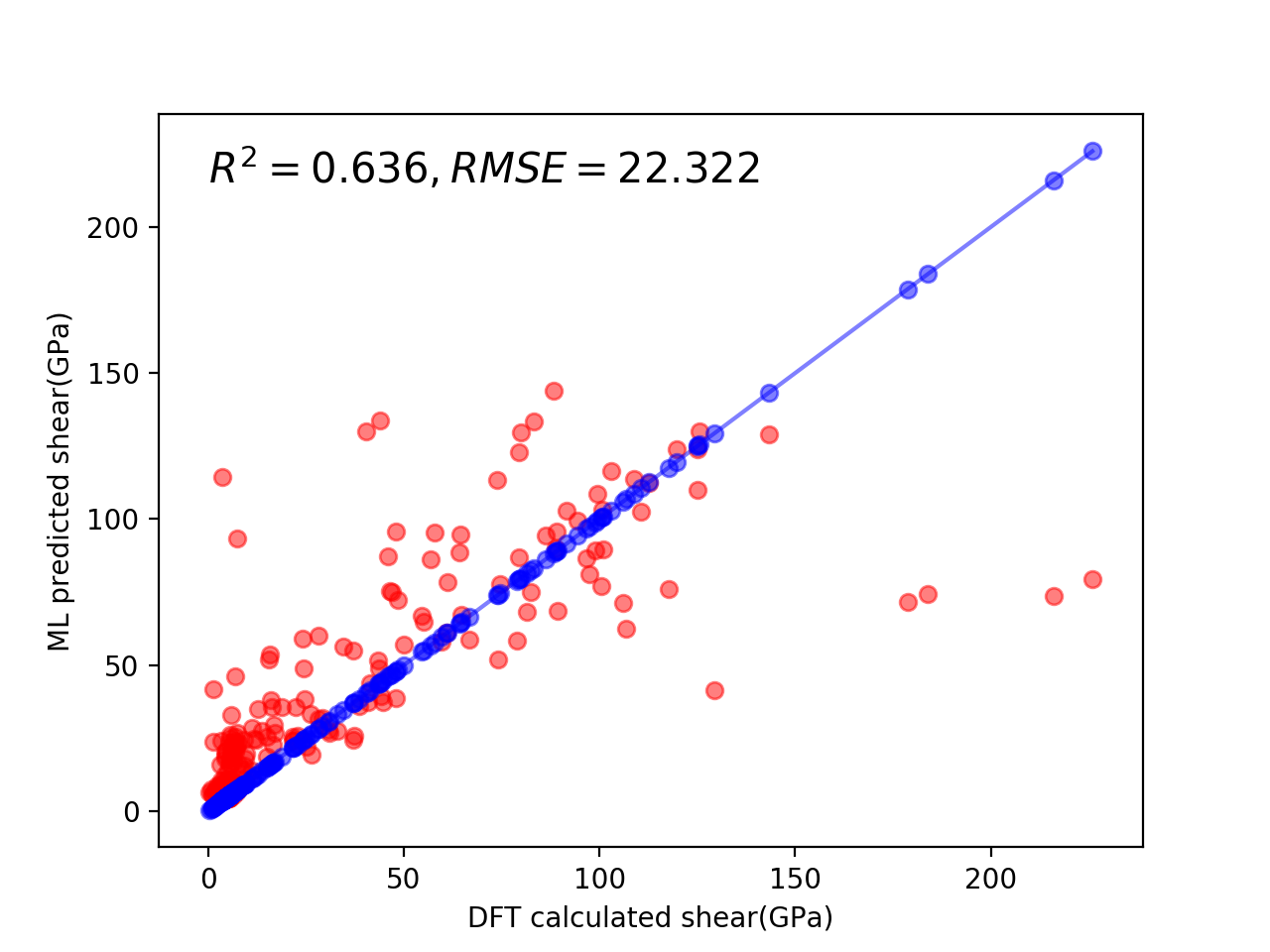}
        \caption[]%
        {{}}
        \label{fig:shear_parity}
    \end{subfigure}

    \caption[]
    {\small Panels (a) and (b) show ML-predicted versus DFT-calculated bulk and shear modulus respectively.} 
    \label{fig:parity_line}
\end{figure}

To further validate our neural network models, we predict the bulk and shear modulus of a set of external materials from the OQMD~\cite{saal2013materials} database and compared them to DFT calculated ones. We first collect all the materials of the space group Fm$\bar{3}$m from OQMD and then remove the duplicates existing in the Material Project database that we used as the training set. We also filter out the materials having more than 40 atoms in the unit cell.  We finally obtain 329 materials as our test set. Then we apply the trained fusion model (Magpie + ECD features) trained with Material Project samples to predict the bulk and shear modulus of the 329 samples in the test set and compared them with DFT-calculated ones as shown in Figure~\ref{fig:parity_line}. We find that our fusion model successfully predicted the bulk modulus for the 329 materials with good alignment with DFT calculated values. The $R^2$ and RMSE in predicting bulk are 0.93 and 21.331 as shown in Figure~\ref{fig:bulk_parity}. However, we also find that the ML-predicted the predicted shear modulus values deviate much more from the DFT calculated ones compared to the bulk modulus, which reflects the fact that it is more difficult to predict shear modulus than bulk modulus. We also observe that most of the deviations of the predicted values compared with DFT calculated ones are from the regions with low bulk or shear modulus and the predicted values usually are large than the DFT calculated ones.

\section{Conclusions}

We propose to combine deep convolutional neural networks and electronic charge density (ECD) for materials elasticity prediction. We demonstrate that the ECD descriptor can be used to predict bulk and shear modulus with CNNs model. 
We created a benchmark dataset named ``FCC2170'' with 2,170 materials of Fm$\bar{3}$m space group from Materials Project database and derived 11 non-redundant leave-one-element-out datasets for benchmarking the proposed ML models with ECD and elemental Magpie features. Our computational experiments showed that due to the structural similarity among the samples of the FCC2170 dataset, the elemental Magpie feature with CNN models achieved the best results, which however, can be enhanced by the fusion models with both Magpie and ECD features. In addition, our benchmark studies on the non-redundant datasets showed that the structure-based ECD feature with CNNs can achieve better extrapolation prediction performances over half prediction tasks out of the total 22 experiments for prediction bulk and shear modulus. 

% and a 3D convolutional neural network (CNNs), that are trained on the dataset to learn condensed features. Our computer experiments show that our method can achieve comparable results to RF with Magpie when applying 5-fold cross validation. Moreover, we conduct extrapolation experiments by removing materials that specifically have one element removed. The results show that our model can achieve much better extrapolation performance than the baseline method in some cases, when it predicts random values for materials. 

To further understand the power of the ECD descriptor, we visualized the distribution of training and test datasets of two descriptor types using t-SNE. It shows that when the training set and testing set of the non-redundant datasets have higher level of mixing, the Magpie-based CNN models work better. When they have lower level of mixing, the ECD descriptor based models significantly outperform the Magpie based CNN models. The results demonstrate the importance of structure based features for achieving higher extrapolation and generalization prediction capability. It is expected that our ECD descriptor with CNN models can also be applied to a variety of problems in material science, especially with the development of algorithms for predicting ECD \cite{gong2019predicting}. Currently, we are generating more ECD dataset with more space groups to extend this method to more materials with diverse structures.

%%%%%%%%%%%%%%%%%%%%%%%%%%%%%%%%%%%%%%%%%%
% \section{Patents}
% This section is not mandatory, but may be added if there are patents resulting from the work reported in this manuscript.

%%%%%%%%%%%%%%%%%%%%%%%%%%%%%%%%%%%%%%%%%%
\vspace{6pt} 

\vspace{6pt} 

%%%%%%%%%%%%%%%%%%%%%%%%%%%%%%%%%%%%%%%%%%
%% optional
%\supplementary{The following are available online at \linksupplementary{s1}, Figure S1: title, Table S1: title, Video S1: title.}

% Only for the journal Methods and Protocols:
% If you wish to submit a video article, please do so with any other supplementary material.
% \supplementary{The following are available at \linksupplementary{s1}, Figure S1: title, Table S1: title, Video S1: title. A supporting video article is available at doi: link.}

%%%%%%%%%%%%%%%%%%%%%%%%%%%%%%%%%%%%%%%%%%
\section{Author contribution}

Conceptualization, M.H. and J.H.; methodology, Y.Z. and J.H.; software, Y.Z.; validation, Y.Z.; data curation, Y.L.; writing--original draft preparation, Y.Z., J.H., and M.H.; writing--review and editing, Y.Z., J.H., M.H., and S.L.; visualization, Y.Z.; supervision, J.H.; project administration, J.H. and M.H.; funding acquisition, J.H. and M.H.

%%%%%%%%%%%%%%%%%%%%%%%%%%%%%%%%%%%%%%%%%%
\section{Funding}

This research was partially funded by NSF under grant number 1940099,1905775,and OIA-1655740 and DOE under grant number DE-SC0020272.

\bibliographystyle{unsrt}
% \bibliography{references}

\end{document}